\documentclass[twocolumn]{revtex4}
\usepackage{graphicx}
\usepackage{dcolumn}
\usepackage{longtable}

\begin{document}

\title
{SU(3) quasidynamical symmetry underlying the Alhassid--Whelan arc of regularity}

\author
{Dennis Bonatsos$^1$,  E.A. McCutchan$^2$, and R.F. Casten$^3$}

\affiliation
{$^1$Institute of Nuclear Physics, N.C.S.R.
``Demokritos'', GR-15310 Aghia Paraskevi, Attiki, Greece}

\affiliation
{$^2$ Physics Division, Argonne National Laboratory,
Argonne, Illinois 60439, USA}

\affiliation
{$^3$ Wright Nuclear Structure Laboratory, Yale
University, New Haven, CT 06520, USA}

\begin{abstract}

The first example of an empirically manifested quasi dynamical symmetry trajectory 
in the interior of the symmetry triangle of the Interacting Boson Approximation model is identified for large boson numbers.
Along this curve, extending from SU(3)  to near the critical line of the first order phase transition,
spectra exhibit nearly the same degeneracies that characterize the low energy levels of SU(3). 
This trajectory also lies close to the Alhassid-Whelan arc of
regularity, the unique interior region of regular behavior connecting the SU(3) and U(5) vertices, thus
offering a possible symmetry-based interpretation of that narrow zone of regularity amidst regions
of more chaotic spectra.

\end{abstract}


\maketitle

Mesoscopic systems are often considered from the complementary perspectives of the degrees of freedom of individual constituents and their interactions, and that of the many-body system as a whole, with its global characteristics, symmetries and quantum numbers.  Studies from the latter perspective can often explain the remarkable regularities such systems exhibit and reveal simple facets, such as collective correlations, that may be difficult to discern from a microscopic approach.

Degeneracies in energy spectra are one manifestation of the symmetries underlying the collective Hamiltonian of many-body systems. For example, degeneracies in the spectrum of the three-dimensional harmonic oscillator reflect the underlying U(3)$\supset$O(3) symmetry
\cite{Smirnov}.  Of course, many-body symmetries are broken in actual physical systems.  The study of symmetry-breaking often reveals subtle, yet important, aspects of the collective physics.  For example, the breaking of the harmonic oscillator degeneracies can shed light on phonon-phonon interactions.

A fascinating recent aspect of symmetry-breaking is that, in some cases, despite considerable deviations of the wave functions from a given symmetry, some properties of that symmetry, such as characteristic degeneracies, persist.  Such systems can be described 
in terms of partial dynamical symmetries \cite{AL25,Lev77,LVI89} [situations for which all of the states preserve part of the 
dynamical symmetry (DS), or part of the states preserve all or part of the DS] 
or quasi dynamical symmetries (QDS) \cite{Rowe1225,Rowe2325,Rowe745,Rowe756,Rowe759} (symmetries that persist in spite of strong symmetry-breaking interactions). In a QDS a significant subset of observables exhibits all the properties
of a symmetry, while others reveal the symmetry to be broken \cite{Rowe745}.

This Letter will focus on a new QDS relevant to atomic nuclei. We work in the context of the Interacting Boson Approximation (IBA) model \cite{IA}, which is couched in a group theoretical framework, with wide applicability \cite{Frank} in molecular and chemical systems. Thus, the ideas developed here should themselves be of wide interest.

The IBA possesses an overall U(6) symmetry with three dynamical symmetries, labelled by subgroups of U(6), namely U(5)
(appropriate for spherical vibrational nuclei), SU(3) (suitable for certain prolate deformed nuclei), and O(6) (proper
for certain axially asymmetric nuclei).  These symmetries are traditionally placed at the vertices of a triangle \cite{Casten}, as in Fig.~1 (top). Certain symmetry aspects persist elsewhere in the triangle. For example, the entire upper left line between U(5) and O(6) contains an underlying symmetry \cite{Talmi} {called O(5)}, which corresponds to potentials independent of the amount of axial asymmetry ($\gamma$-unstable potentials). Along this line there is a high degree of energy degeneracy. QDS occur along other legs of the triangle \cite{Rowe1225,Rowe2325,Rowe745,Rowe756,Rowe759} in which certain properties of U(5), SU(3), or O(6) are preserved (except near the phase transitional points). No other line possessing any underlying symmetry higher than SO(3), necessary for rotational invariance and present everywhere in the triangle, is known to exist.

\begin{figure}
\center{{\includegraphics[height=95mm]{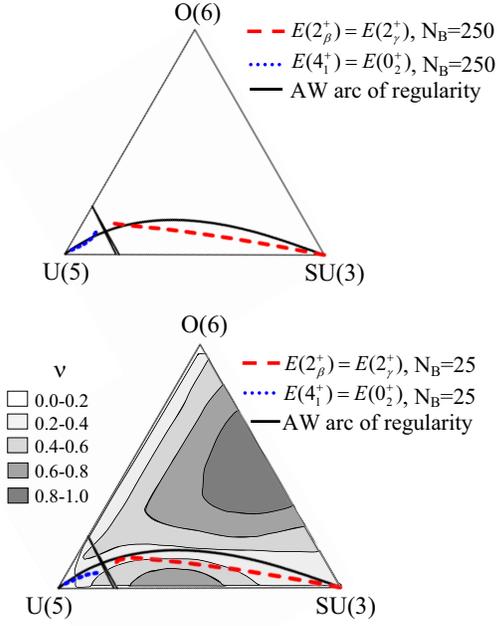}}}
\caption{(Color online) IBA symmetry triangle in the parametrization of Eq. (\ref{HAW})
with the three dynamical symmetries and the Alhassid--Whelan arc of regularity [Eq. (\ref{arc})]. 
The shape coexistence region \cite{IZC} between spherical and deformed phases, shown by slanted lines
near the U(5) vertex, encloses a first order phase transition terminating in a point of second order transition on the U(5)-O(6) leg.
The loci of the degeneracies $E(2_\beta^+)$=$E(2_\gamma^+)$ (dashed line on the right, corresponding to the QDS discussed in this Letter)
and  $E(4_1^+)$=$E(0_2^+)$ (dotted line on the left) are shown for $N_B$=250 (top) and
$N_B$ = 25 (bottom).  In the bottom part, the $\nu$-diagram, based on Ref. \cite{AWNPA}
is shown.}
\end{figure}

It is the purpose of the present Letter to show the first example of a QDS trajectory in the $interior$ of the triangle. It is based on SU(3) and its locus, shown in Fig.~1, extends from SU(3) nearly to the region of coexistence \cite{IZC} of spherical and deformed phases. Moreover, we will show that this curve lies within a narrow internal region proposed twenty years ago \cite{AWPRL,AWNPA}, called the arc of regularity, in which spectra exhibit a high degree of order, whereas chaotic behavior is found nearly everywhere else. It has long been speculated that this arc (solid curve in Fig.~1) reflected some underlying symmetry
but the nature of that symmetry had not heretofore been elucidated.  
We suggest, from an extensive analysis of the low energy part of the 
spectrum, that that symmetry is, in fact, the SU(3) QDS we have identified.

In what follows we use the IBA Hamiltonian \cite{AWPRL,AWNPA,AWexp}
\begin{equation}\label{HAW}
H(\eta,\chi) =
c \left[ \eta \hat n_d +{\eta- 1 \over N_B}
\hat Q^\chi \cdot \hat Q^\chi\right],
\end{equation}
where $\hat n_d = d^\dagger \cdot \tilde d$, $\hat Q^\chi =
(s^\dagger \tilde d + d^\dagger s) +\chi (d^\dagger \tilde
d)^{(2)},$ $N_B$ is the number of valence bosons, and $c$ is a
scaling factor. This popular Hamiltonian contains two parameters,
$\eta$ and $\chi$, with $\eta$ ranging from 0 to
1, and $\chi$ ranging from 0 to $-\sqrt{7}/2$=$-1.32$ and allows for symmetry breaking. Any pair of ($\eta,\chi$) values can be mapped \cite{McC69} onto the triangle.
For $\chi$=$-\sqrt{7}/2$ the shape coexistence region occurs at $\eta \sim$ 0.8.  Eq. (\ref{HAW}) has
an alternate form using parameters ($\zeta$, $\chi$).  The coexistence region then occurs for $\zeta \sim$ 0.5 but the physics is identical.
We utilize the $\eta$ form to be consistent with Refs.~\cite{AWPRL,AWNPA}. Numerical calculations of energy levels have been performed using the code IBAR \cite{IBAR,IBAR2}, which can handle boson numbers up to $N_B$=250.

A hallmark of SU(3) is degeneracies within sets of bands 
comprising a given irreducible representation (irrep), such as those between the levels of the $\beta$ band
and those (with the same [even] $L$) of the $\gamma$ band. To inspect whether these degeneracies approximately persist for parameters that deviate from SU(3) we use, as a measure of degeneracy breaking between the $\beta$ and $\gamma$ band, the rms deviation
\begin{equation}\label{sigmabg}
\sigma_{\beta\gamma}=\sqrt{ \sum_2^{L_{max}} [E(L_\beta^+)-E(L_\gamma^+)]^2 \over {L_{max}\over 2}-1 },
\end{equation}
where $L_{\beta}^+$=$L_{\gamma}^+$ and with all energies normalized to $E(2_1^+)$.

Another hallmark of SU(3) behavior at low energies
is the position and degeneracies of the $0^+$ bandheads belonging to different irreps, as 
determined by the eigenvalues of the second order Casimir operator of SU(3) and reported in Ref. \cite{PRL101}. In order to examine
to which degree the $0^+$ states occurring in a calculation obey the SU(3) rules,
we use the relevant rms deviation
\begin{equation}\label{sigma0}
\sigma_{0}=\sqrt{ \sum_3^{i_{max}} [E(0_i^+)^{th}-E(0_i^+)^{\rm SU(3)}]^2 \over i_{max}-3 }.
\end{equation}
\noindent with all energies normalized to $E$($0_2^+$) and considering the lowest nine 
0$^+$ states (i.e., $i_{max}$=9).

\begin{figure}
\center{{\includegraphics[height=120mm]{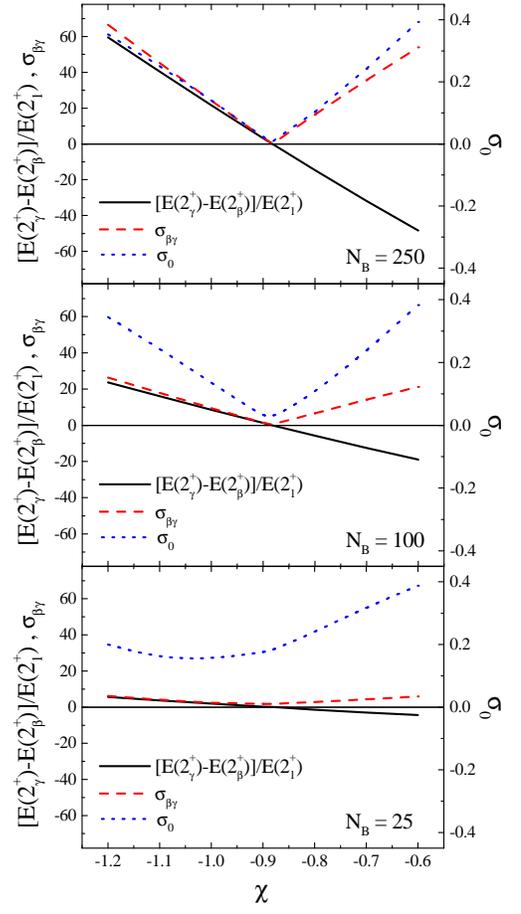}}} \caption{
(Color online) The energy difference $E(2_\gamma^+)-E(2_\beta^+)$ (normalized to $E(2_1^+)$)  and
the quality measures $\sigma_{\beta\gamma}$ [Eq. (\ref{sigmabg}), up to $L_{max}$=10] and  $\sigma_0$
[Eq. (\ref{sigma0}), up to $i_{max}$=9], are shown for $\eta$=0.632, varying $\chi$, and
boson numbers $N_B$=25, 100, 250.}
\end{figure}

\begin{figure}
\center{{\includegraphics[height=50mm]{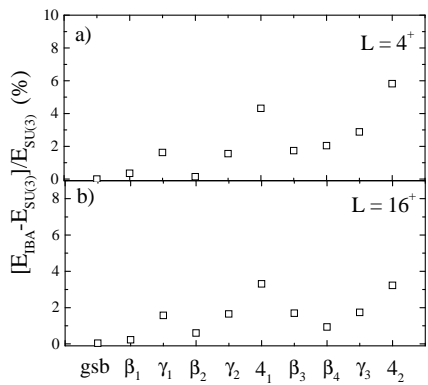}}}
\caption{Percent difference between SU(3) predictions and the IBA results [$N_B$=250, $\eta$ = 0.632, $\chi$=-0.882]
for (a) $L$=$4^+$ and (b) $L$=$16^+$ states.  The corresponding bandhead names are given on the horizontal axis.}
\end{figure}

\begin{figure}
\center{{\includegraphics[height=80mm]{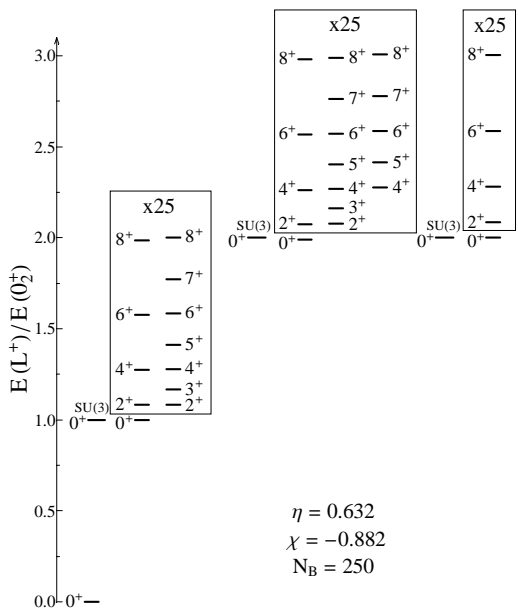}}}
\caption{Level scheme for an IBA calculation at a point where $E(2^+_\beta$)=$E(2^+_\gamma$). 
An expanded energy scale (x25) is used within the boxes to show the small rotational energies.
The SU(3) $0^+$ bandheads are also shown.}
\end{figure}

The tools used in this study ($\sigma_{\beta\gamma}$ and $\sigma_0$) are based on the degeneracies and regularities 
exhibited by SU(3) for large boson numbers and at low energy. At higher energies the 3D harmonic oscillator degeneracies of $0^+$ states
fade away (for $N_B=250$ these degeneracies are exhibited clearly by the lowest 110 $0^+$ states out of 5334), 
thereby also removing degeneracies between higher $L$ states built on different $0^+$ bandheads. For these higher energies, statistical tools 
would have to be used \cite{AWPRL,AWNPA}.

We start by looking for the point of the $E(2_\beta^+)$=$E(2_\gamma^+)$ degeneracy by keeping $\eta$ constant and varying $\chi$ (a trajectory parallel to the O(6)-SU(3) line but interior in the triangle for $\eta \ne 0$). We find, as illustrated in Fig.~2 for $\eta$=0.632, that the location of $E(2_{\beta}^+)$=$E(2_{\gamma}^+)$ also corresponds to the minima in  $\sigma_{\beta\gamma}$ and $\sigma_{0}$.  This means that when $E(2_{\beta}^+)$=$E(2_{\gamma}^+)$ one simultaneously has $E(L^+_\beta)\sim E(L^+_{\gamma})$ for all even $L$ values, the same degeneracies that characterize SU(3). An even more significant result is that, at the point where $E(2^+_\beta)$=$E(2^+_\gamma)$, not only the $\beta$ and $\gamma$ bands, 
but $all$ low-lying bands, to an excellent approximation, have the same degeneracies as in SU(3).
In addition, all bands possess practically the same moments of inertia, 
following the $L(L+1)$ dependence on angular momentum quite accurately. 
Figure 3 illustrates the quality of these degeneracies. For nearly all states, the deviation from SU(3) is $<$5$\%$.

Perhaps more surprising, the simultaneous minimum in $\sigma_{0}$ means that the $0^+$ bandheads 
also possess almost the same relative energies as in SU(3). That is, even though energies of sets of bands 
in a given SU(3) irrep may have moved substantially, ratios of bandhead energies remain nearly as in SU(3) 
and, $within$ each set, the full set of SU(3) energy degeneracies persists (see examples in Fig. 4).  
The minima are sharper for increasing $N_B$. $N_B$=25 is relevant for actual  nuclei and 
$N_B$=250 approaches the classical limit. 

Moreover, at the point of $E(2_\beta^+)$=$E(2_\gamma^+)$, intraband $B(E2)$ ratios within  the ground and $\beta_1$ bands
exhibit SU(3) values, while interband $B(E2)$s [normalized to $B(E2; 2_1^+\to 0_1^+)$] belonging to different SU(3) irreps,
which vanish in the SU(3) limit, remain at least three orders of magnitude lower than intraband $B(E2)$ values.

\begin{figure}
\center{{\includegraphics[height=80mm]{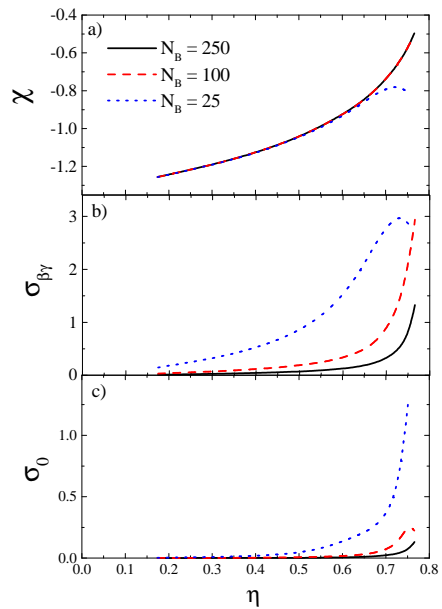}}}
\caption{(Color online) The $\vert \chi \vert$ parameter values providing the degeneracy $E(2_{\beta}^+)$=$E(2_{\gamma}^+)$
and the quality measures $\sigma_{\beta\gamma}$ [Eq. (\ref{sigmabg}), up to $L_{max}$=10]
and  $\sigma_0$ [Eq. (\ref{sigma0}), up to $i_{max}$=9], are shown for different values
of  $\eta$ and  $N_B$=25, 100, 250.}
\end{figure}

Figures 2--4 gave results for $\eta$=0.632. However, the Hamiltonian actually gives a locus of ($\eta, \chi$) points where the condition $E(2^+_\beta)$=$E(2^+_\gamma)$ is met (dashed line to the right of the transition region in Fig.~1). For $N_B$=250, it ranges from ($\eta, \chi$) $\sim$ (0.8, -0.5) to (0, -1.32) at SU(3), with only a weak dependence on $N_B$ except near the coexistence region, as seen in Fig. 5(a).
Figures 5(b) and (c) show the SU(3) degeneracy measures $\sigma_{\beta\gamma}$ and $\sigma_0$ along this locus. They are extremely small 
from SU(3) ($\eta$=0.0) to $\eta$ values very close to the coexistence region ($\eta$=0.8).  For example, for $N_B$=250, 
$\sigma_0$ $<$0.1 almost to $\eta$=0.8 and $\sigma_{\beta\gamma}$ $<$ 0.4 down to $\eta$ $\sim$ 0.7.  
Near SU(3), the quality of the SU(3) degeneracies depends only slightly on $N_B$, but drops substantially ($i.e.$, $\sigma_{\beta\gamma}$ and $\sigma_0$ increase) for lower $N_B$ near the coexistence region.

The above findings suggest that (for large $N_B$) fundamental aspects of SU(3) 
at low energies---degeneracies, relations among the 0$^+$ bandheads, 
and $B(E$2) values---are robust, not limited to its vertex, but extending 
deeply into the triangle, until the critical line is approached. 
The degeneracies within irreps, and among different 
irreps at low energy in SU(3), are characteristic signatures of an SU(3) QDS 
and thus the locus of these degeneracies is the first known example of a 
QDS curve in the interior of the symmetry triangle.
Such degeneracies persist though the wave functions are strong 
admixtures of SU(3) basis states.  This is reminiscent of effects of 
small perturbations of nearly degenerate levels that result in strong 
mixing but accompanied by small energy shifts \cite{Hecht}.

Given that QDS involve degeneracies (regularities in energies), it is not surprising that their region is one of higher overall regularity.
Alhassid and Whelan \cite{AWPRL,AWNPA}
discovered a unique arc of regularity which starts at SU(3) and curves inside the triangle to the U(5) vertex (see Fig. 1), approximately described \cite{CJPRE} by
\begin{equation}\label{arc}
\chi(\eta)= {\sqrt{7}-1 \over 2} \eta - {\sqrt{7} \over 2}
\end{equation}
for $N_B$=25 and depending only weakly on $N_B$ \cite{AWNPA}. In Ref.~\cite{AWexp} 12 nuclei were shown to lie close to the arc.

It has long been suspected that the arc reflected an underlying symmetry but none has ever been found.  However, the locus of SU(3) degeneracies in the triangle, defining the SU(3) QDS discussed above, corresponds closely to that of the arc of regularity [see Fig.~1 (top)], suggesting that this interior QDS may provide its underlying symmetry.
This is also evident in the $\nu$-diagram in Fig.~1 (bottom) [$\nu$ is a measure of chaos, low values of $\nu$ indicating regularity].
Thus, the present results show not only the first example of an interior QDS line, but also that it closely tracks the unique example of an interior arc of regularity.
[Degeneracies along the line of first order phase transition have recently raised suspicions \cite{PRL100} of another interior QDS, but the problem remains open.]

We note that, using the intrinsic state formalism in the $N_B \to \infty$ limit, Ref.~\cite{Macek} found that 
the degeneracy of the $\beta_1$ and $\gamma_1$ bandheads
practically coincides with the present line of $E(2_\beta^+)$=$E(2_\gamma^+)$ for $N_B$ = 250,
as well as to the locus of changes in properties of the $\gamma$-vibration. 
In addition, the region of validity of the SU(3) QDS inside the Casten triangle, as well as critical behavior 
within the low-lying spectrum due to a degeneracy of the $\beta$- and $\gamma$ vibrations, have been considered in detail in Ref.
\cite{Macek2}. 

To the left of the coexistence region, one can consider the
$E(4_1^+)$=$E(0_2^+)$ degeneracy, which characterizes U(5) \cite{IA}. This line is included in Fig.~1. However, for $N_B=250$, the fingerprints of U(5) are actually well preserved in the whole region between U(5) and the critical line, and
no sharp minima of the type in Fig.~2 are found for U(5) degeneracy measures.
This agrees with Fig.~1 (bottom), where most of the region between U(5) and the critical line exhibits a high degree of
regularity (low values of $\nu$).

In conclusion, using the degeneracy $E(2_\beta^+)$=$E(2_\gamma^+)$, we have discovered a trajectory (dashed line in Fig. 1) from 
SU(3) to near the phase coexistence region, along which spectra exhibit nearly the same degeneracies 
as exhibited by SU(3) at low energies for large boson numbers, 
thus offering the first example of a quasidynamical symmetry curve in the interior of the triangle.
The SU(3) degeneracies deteriorate with decreasing boson number, but the locus remains almost invariant. The trajectory of the QDS falls within the Alhassid--Whelan arc of regularity, the unique region of regular behavior connecting SU(3) and U(5). These results suggest that there is indeed a symmetry underlying the arc of regularity, at least at low energies, and that it is the SU(3) QDS.

The authors are grateful to R.~J. Casperson for the code IBAR, which made the present study possible,
to S. Heinze for an IBA code for B(E2)s, and to V.~Werner, E.~Williams, J. Dobe\v{s}, P. Cejnar, 
and M. Macek for useful discussions.
Work supported by U.S.~DOE Grant No.~DE-FG02-91ER-40609 and by the DOE Office of Nuclear Physics under Contract DE-AC02-06CH11357.

\end{document}